\documentclass{elsart}

\usepackage{natbib}
\usepackage{graphics,graphicx,amssymb}

\begin{document}

\begin{frontmatter}

\title{Cosmic dust optical properties:
numerical simulations and future laboratory measurements in microgravity}
\author[SA_UPMC]{J. Lasue}
\author[SA_UPMC]{A.C. Levasseur-Regourd}

\address[SA_UPMC]{Service d'A\'eronomie et Universit\'e Pierre et Marie Curie, 
Route des G\^atines, BP3-91371 Verri\`eres (France).}

\begin{abstract}
Understanding the properties of particle aggregation and resulting aggregates under microgravity 
conditions leads to better insights on the formation of the early Solar System planetesimal. 
Simulating such conditions is the main objective of Interactions in Cosmic and 
Atmospheric Particle System (ICAPS), a multi-users facility currently under phase B at ESA for the 
International Space Station (ISS). 
First results of light scattering simulations
by core-mantle aggregates of grains with organics and icy mantles are presented to show the evolution
of polarization with aggregation. 
The Light Scattering Unit (LSU) is both a polarization diagnostic tool for ICAPS, and an experiment that 
will allow the interpretation of the available 
light scattering dust observations in terms of physical properties of the scattering media. 
This presentation updates the current approach of the calibration procedures and the innovative 
experimental setup (providing a phase angle exploration from about $2^{\circ}$ to $175^{\circ}$ 
together with 
a wavelength exploration from 0.4 to 0.8 micrometers). We also assess the possibility for 
the determination of the entire Stokes vector of the light scattered by the aggregates. 
\end{abstract}

\begin{keyword}
polarization \sep light scattering \sep cosmic dust \sep microgravity experiment.

\end{keyword}

\end{frontmatter}

\section{Introduction}
In the proto-solar nebula, solid submicron sized grains from interstellar origin 
slowly merged to form solid particles, small bodies and planets~\citep{jmg_rev97}. 
In the Solar System, cosmic dust can be found in dust clouds and cometary comae, or can form regolith 
on  asteroidal surfaces and aerosols in planetary atmospheres.
Physical characteristics about these particles come from in situ studies by 
space missions such as Stardust, see e.g.~\citet{aclr_sci04,ajt_tee04}, 
and collection of Interplanetary Dust Particles (IDPs) collected in the 
earth atmosphere, see~\citet{ez_tlw94,ekj_ts01}.
Nevertheless, most of the available data regarding the physical properties of 
the particles (size, morphology, albedo etc.) are deduced from remote observations of 
the light scattered by the dust. 

             \subsection{Observations of scattered light}

The complete description of scattered light is given by a four elements vector, also called Stokes vector,
the coefficients of which are  $I,Q,U,$ and $V$, respectively the total intensity of 
the light, the difference of parallel ($I_{\parallel}$) and perpendicular ($I_{\perp}$) 
polarized intensity with respect to the scattering plane, 
the difference of the intensities of linearly polarized light at $+45^{\circ}$ and 
$-45^{\circ}$ and finally the difference between left and right polarized intensity~\citep{bh93}.
Observations of the linear polarization degree, $P$, defined as: 
\begin{equation}
P=\frac{I_{\perp}-I_{\parallel}}{I_{\perp}+I_{\parallel}}=-\frac{Q}{I}
\end{equation}
are of special interest.
Being a dimensionless number and the incident solar light being unpolarized, 
$P$ does not require any calibration with the distances
between the scatterer and the source or observer, 
but only varies with the phase angle $\alpha$, the wavelength $\lambda$ and the dust physical properties.

Current polarimetric observations show that most of the  dust clouds, cometary comae 
and asteroidal regoliths in the Solar System present similar linear polarization 
phase curves. They all have a bell-like shape with a small negative branch (electric field parallel to the 
scattering plane) for small phase angles, an inversion angle around 20$^\circ$ and a large positive 
branch with a maximum value in the [80$^\circ$,120$^\circ$] phase angle range~\citep{msh03,aclr_yal04}. 
The global shape of these curves is most probably 
related to the interaction of light with irregular particles of size larger than the 
wavelength~\citep{cb_dh83,mim_jwh99}.

Analysis of recent observations have shown a quasi-linear dependence of the polarization with 
respect to the wavelength in the visible domain. The gradient is positive for 
 cometary comae, and negative for  S-type asteroids. 
In both cases, the slope increases with larger phase angles between the inversion and the maximum 
region~\citep{aclr_eh03}.

\subsection{Light scattering measurements on aggregates}

The IDPs constituted by aggregates of submicron grains are generally expected to be from cometary origin~\citep{fr02}. 
Such particles exhibit very low densities as reviewed by~\citet{fr98}, in agreement with the density
of about 100 $kg.m^{-3}$ obtained from the Giotto probe for Halley's comet~\citep{mf_aclr00}.
The aggregation of interstellar grains into very fluffy particles was suggested by~\citet{jmg_jih90} 
to explain the formation of cometary grains and cometary nuclei.

Experiments on light scattering by particles have been developed by laboratory to reproduce 
the intensity and polarimetric curves observed in the Solar System. 
A couple of experiments have been performed with slowly falling 
particles, including dust from meteorites, thus validating the experimental approach and presenting 
relevant results regarding the polarization phase curve~\citep{raw_lrd97,om_hv00,eh_jbr03}. 
Other relevant experiments made use of the similarity principle and microwave techniques
with bigger particles~\citep{basg_lk99}. Even if interesting results have been obtained, 
ground-based experiments cannot entirely reproduce the space environment 
in which dust particle aggregation processes took place and we need other approaches 
to improve our knowledge of the dust particle properties.

In the next part, we will describe results obtained through a numerical simulation 
of cometary dust particles. Light scattering codes are efficient and can be used for  an 
increasing number of particle types. However the computer time required for such applications 
is still very high. Moreover the approximations used in the algorithms should be checked with 
 well controlled laboratory measurements.

Microgravity experiments allow us to create experimental conditions closer to the space environment
in which cometary and interplanetary dust lie. Therefore particle aggregation can be reproduced 
free from gravitational forces. The drawback of working under space conditions is that the experiments 
have serious constrains in terms of size, mass and time of duration.  
More precisely, the light scattering measurements that will be performed with ICAPS, a multi-users 
microgravity facility for the ISS, will be further detailed in the last part of the paper.

\section{Core-mantle model for cometary dust particles}

\subsection{Description of the model}

The description of light scattering by irregular particles with size close to the incident 
wavelength involves numerical tools and approximations, such as 
the Discrete Dipole Approximation (DDA)~\citep{btd_pjf00}, or the T-matrix~\citep{mim_jwh99}.
These tools have already been used for the description of cometary particle 
properties using aggregates of spheres, showing the apparition of a negative branch 
and a maximum in polarization at $90^{\circ}$
for porous aggregates of few tens of submicron spheres, see e.g.~\citet{zx_mh97,evp_kj00}.

One  commonly accepted model used to explain the astronomical observations 
and physical properties of pre-cometary grains, or proto-solar grains, is the core-mantle 
model presented by~\citet{jmg_jih90}. The interstellar grains are 
elongated and constituted of a core of interstellar silicates,
 a layer of organic refractory material 
and an outer layer of volatile ices with water ice as the main component.

As a first step, we use spherical grains,
with two concentric spherical layers:  a silicate core
with refractive index $m=1.62+0.003\times i$ \citep{jd_bb95}, 
a first layer of organic material with $m=1.88+0.1 \times i$ \citep{jmg_al96}
and a second layer of water ice, the refractive 
index of which might be assumed to be $m=1.31$ \citep{sgw84}. 
All the indices are given in the optical domain.
The mean size parameter of the grains, $X=\frac{2\pi r_{grains}}{\lambda}$,
is around 2, at $\lambda =0.5$ $\mu m$, as deduced from~\citet{li_jmg97}.

\subsection{First results}

We used the DDA code to calculate the light scattered 
by  such grains aggregated by a typical fractal process: Ballistic Cluster-Cluster 
Aggregation (BCCA),  described by~\citet{pm83}.
When each grain has a size parameter of 2, typical features of the polarization and intensity 
curves observed in the Solar System begin to appear. Fig. 1
shows the evolution of the polarization phase curve expected 
during the aggregation process by comparing in Fig. 1a the polarization phase curve 
for a single two-layered sphere ($X=2$), and in Fig. 1b the curve for a BCCA aggregate of 64 spheres
($X=1.8$ to $2$). The negative branch appears below $20^{\circ}$ phase angle and 
the large positive branch has a maximum around 80$^\circ$. 
Moreover, calculations made for aggregates of spheroidal grains show the same tendencies, the gradient of the 
polarization degree with the size parameter being in this case less steep than in the case of spheres.

\begin{figure}[!h]
\center{\includegraphics[width=2.7in,height=2in]{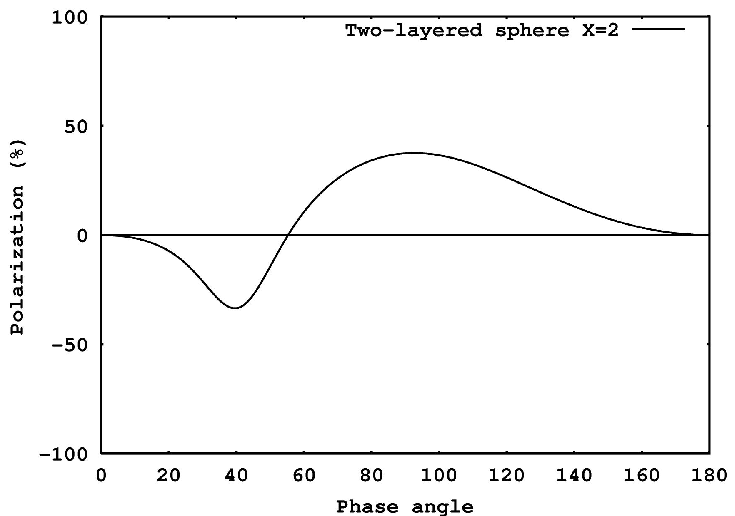} 
\includegraphics[width=2.7in,height=2in]{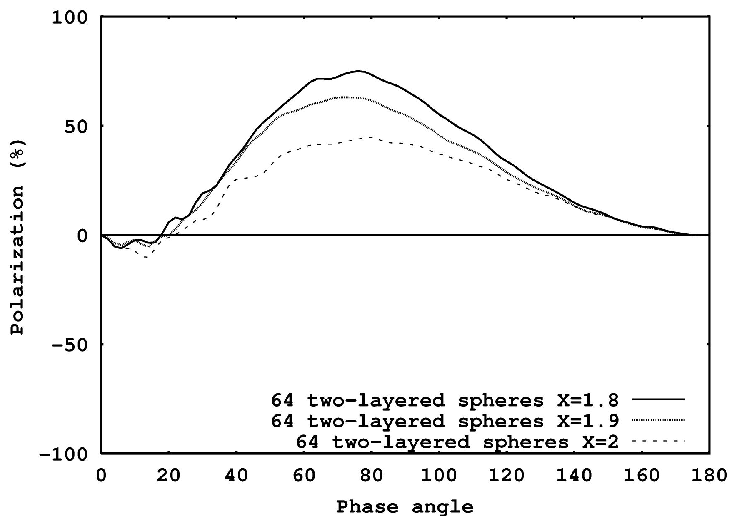}}
\caption{Variation of polarization phase curve with the aggregation: Fig. 1a, Mie theory for a 
two-layered sphere: silicates-organics-ice (X=2); Fig. 1b, 
DDA calculations for a BCCA aggregate of 64 such two-layered spheres (X=1.8; X=1.9; X=2).}
\end{figure}

Our model size, composition and optical indices, even with the high 
simplicity of a spherical hypothesis, produce curves presenting features observed for dust in
the Solar System. 
However, the validity of any numerical approach needs to be confirmed experimentally
by carefully controlling each parameter involved in the model.

\section{ICAPS light scattering description}

The above mentioned process of BCCA
has already been monitored during microgravity experiments 
attempting to simulate the original seedling of planets, producing open structures of 
particles with a fractal dimension around 2~\citep{jb_jd00}.
In order to  understand the properties of particle aggregation and resulting aggregates 
under low-pressure and microgravity conditions,  measurements are performed during microgravity flights 
on a cloud of particles, to analyze their polarization properties and to 
provide a link between the polarization observations and their physical properties.
Reconstructing such realistic conditions is the main 
objective of Interactions in Cosmic and Atmospheric Particle System (ICAPS), 
a multi-users facility currently under phase B at ESA for the ISS.

\subsection{Precursor Experiments}
Precursor experiments have already been performed.
First the PRopri\'et\'es Optiques des GRains Astronomiques et Atmosph\'eriques
(PROGRA$^2$) experiment has shown the feasibility 
of light scattering measurements under microgravity conditions during parabolic flights.
It also allowed the creation of a polarization database for more than 100 samples
including compact as well as fluffy particles~\citep{jcw_jbr99,eh_jbr02}.
Second, the COsmic Dust AGgregates - Sounding Rocket Experiment (CODAG-SRE) 
 has demonstrated the feasibility of simultaneous measurements
 of the polarization degree and total intensity at different phase angles.
The onset of particle aggregation has been monitored, up to aggregates of a few tens of grains~\citep{aclr_vha01}.
New concepts have been developed, which are to be validated 
and will provide new scientific results during the ICAPS Precursor Experiment (IPE) 
due to fly on board the ISS in the near future.

\subsection{Scientific objectives of ICAPS}

The main scientific objectives of the ICAPS experiment are amongst other ones~\citep{jb_mc99b}:
\begin{enumerate}
\item{to simulate the aggregation processes of submicron and micron spheres of several 
absorbent and nonabsorbent materials under presumed proto-solar nebula conditions,}
\item{to produce large aggregates of such spheres whose properties are comparable to 
the ones of cosmic dust and asteroidal regolith,}
\item{to relate the light scattering measurements with the  physical properties 
of the aggregates and regolith simulant formed in the chamber,} 
\item{Finally the scattering measurements shall also validate the approximations used on 
light scattering codes and models.}
\end{enumerate}

The ICAPS light scattering device will be able to probe the polarization degree and total intensity for small phase
angle (as small as $2^{\circ}$) and give information related to the formation of large ($> 1$ $cm$) aggregates. 
The use of three different wavelengths well distributed in the 
visible will allow the monitoring of the linear dependence of the polarization with the wavelength.
Finally, the four Stokes parameters 
(I,Q,U,V) of the scattered light should be retrieved for one phase angle, giving twice as much information as would be 
obtained using a single separating prism. Of particular interest will be the detection of the circular
polarization degree for comparison with the astronomical observations and ground-based 
experiments~\citep{jwh_hv03}.

\subsection{Description of the instrument}

\begin{figure}[!h]
\center{ \includegraphics[width=2.7in,height=2in]{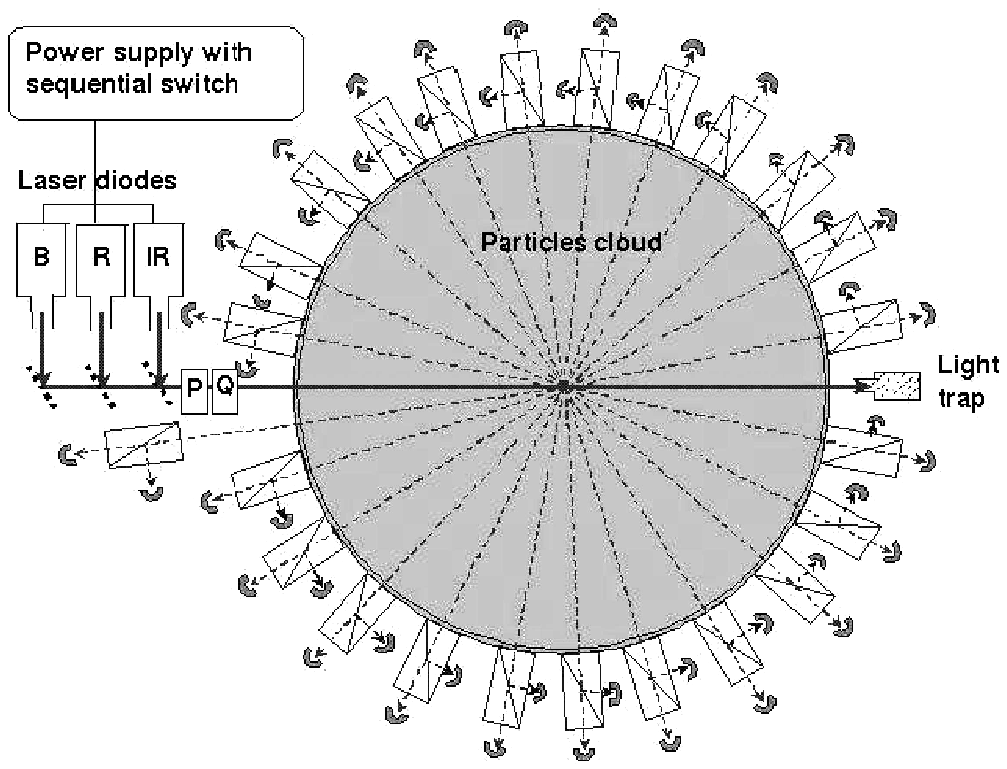}
\includegraphics[width=2.7in,height=2in]{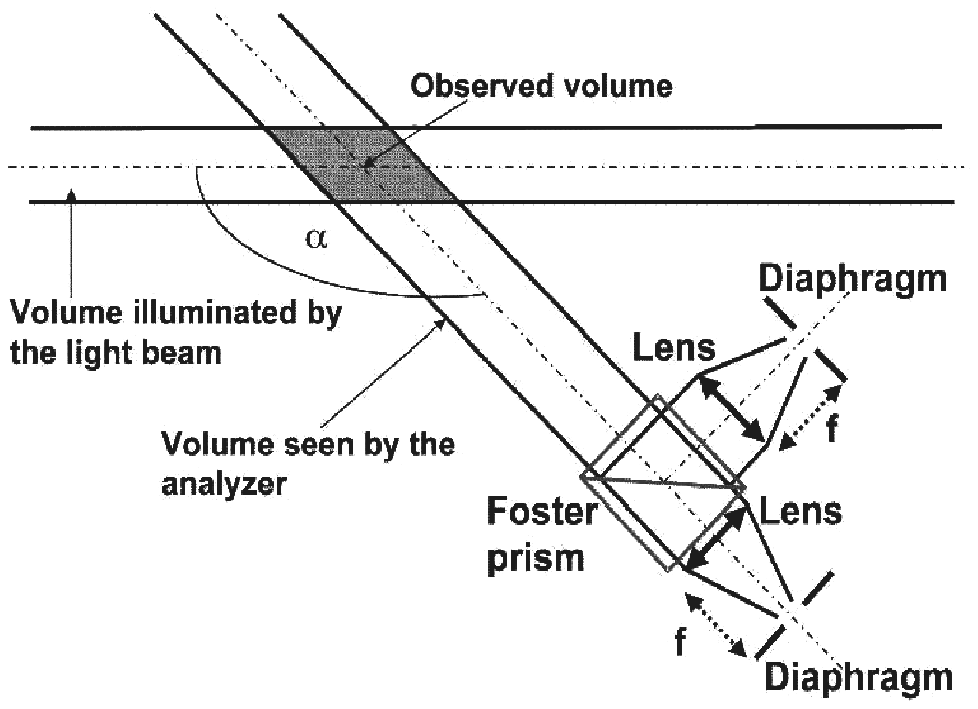}}
\caption{Instrumental concept: left, low pressure chamber with dust particles sequentially illuminated 
by 3 laser diodes; $I$ and $P$ analyzed at 24 phase angles to retrieve phase curve; right, detail of one 
analyzer constituted of a Foster prism with 2 lenses and diaphragms.}
\label{vol_mes_ring}
\end{figure}

In order to investigate the evolution in shapes and characteristics of the
aggregates formed in the dust chamber (Fig.~\ref{vol_mes_ring}), the particles 
will be illuminated by laser diodes and flash lamps
and observed through cameras, microscopes and a new light scattering device upgraded
from the CODAG-LSU. 
Angular light scattering measurements are usually performed by nephelometers, in which a collimated 
light source illuminates a sample and a detector collects the scattered light at a given phase 
angle~\citep{cb_dh83}. 
The sample inside the chamber will be illuminated at three wavelengths (408, 635 and 830 $nm$)
with a high frequency switch (about 100 Hz). 
The unpolarized state of the solar light cannot be reproduced 
satisfactorily by a laser diode. However since we will 
 retrieve the linear polarization degree, the polarization of the incident light on the particles will be previously 
circularized by a Fresnel prism (equivalent to an achromatic quarter-wave plate) in order to give the same intensity 
to the perpendicular and the parallel linear components of the light.

The two perpendicular components of the light scattered by the particles
 are then separated using a Foster prism. 
Each intensity is measured with a very sensitive photodiode
for at least 24 phase angles at three different wavelengths (see Fig.~\ref{vol_mes_ring}).
The specificity of the experiment relies on simultaneous measurements of the polarization
for each phase angle with an array of fixed analyzers, in a small and compact configuration. 
The unscattered and forward scattered light will be absorbed by 
a light trap to prevent stray light from coming back into the chamber.

\begin{figure}[!h]
\center{\includegraphics[width=2.7in,height=1.7in]{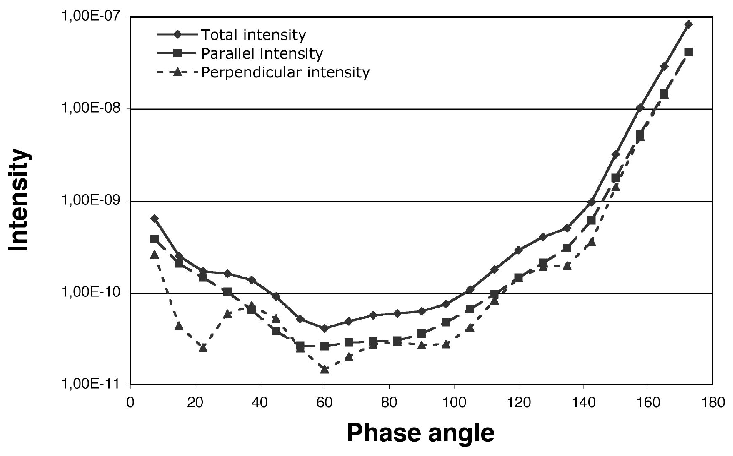} 
\includegraphics[width=2.7in,height=1.7in]{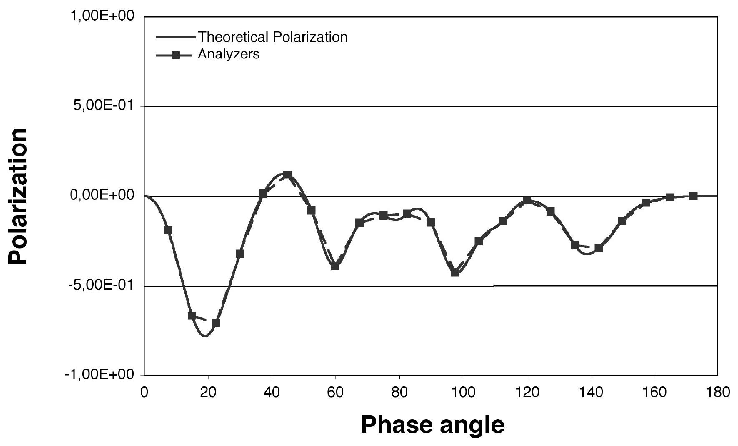}}
\caption{Example of curves the LSU will retrieve from samples of spheres at the beginning
of the aggregation process. Left, intensity and right, polarization  at 650 $nm$ for 0.5 $\mu m$
radius spheres.}
\label{int_pol_650}
\end{figure}

To ensure a proper calibration of the instrument, each analyzer has to be checked separately. 
Tests will be carried out with opal glasses and lambertian surfaces with a well known scattering function. 
A further test using silica spheres on  microgravity (parabolic flight) will  provide 
the definite calibration test and validate the experiment concept with the Mie theory. 
Fig.~\ref{int_pol_650} shows the curve that would be retrieved
by the light scattering unit at 650 $nm$ for spheres of 0.5 $\mu m$ radius.

\section{Preliminary results on the instrument capabilities}

\subsection{Validation of the optical principles}

The breadboard activities of the ICAPS optical concept validated and characterized critical optical elements. 
The Fresnel and Foster prisms  have been studied and proved to be
sufficiently achromatic for the purpose of the experiment. Precision of circularization for 
the Fresnel prism is about 1\% and the separation precision of the Foster prism for each ray of light is 
better than  $\frac{1}{1000}$. Absorption of each of those two elements is lower than 10\%.
The photodiodes have  a high enough sensitivity to detect $3.10^{-11} W$ which is the lowest expected intensity 
on the detectors  based on the physical parameters~\citep{vh00}.

\subsection{First steps towards Stokes polarimetry}

\begin{figure}[!h]
\center{\includegraphics[width=2.7in,height=1.3in]{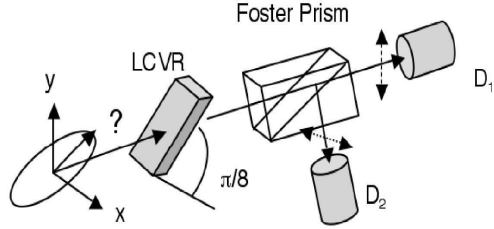}}
\caption{Principle of Stokes polarimetric measurement using a LCVR aligned with a Foster prism.}
\label{Stokes1}
\end{figure}

In order to measure the four Stokes parameters with limited volume, weight and power consumption, 
a Liquid Crystal Variable Retarder (LCVR) aligned with a Foster prism can be used as shown in Fig.~\ref{Stokes1}.
Calculations show that with an alignment angle of $\frac{\pi}{8}$ between the LCVR and the Foster prism, 
the Stokes parameters of the light and the measured intensities on the detectors D, $I_n$, 
are related with the formulae:


\begin{equation}
\begin{array}{l}
I  =I_1+I_2  =I_3+I_4  =I_5+I_6 \\
Q  =I_1-I_2 \\
U  =I_5-I_6 \\
V  =\sqrt{2}\times (I_5+I_1-2I_3)  =\sqrt{2}\times (2I_4-I_2-I_6) \\
\end{array}
\end{equation}

First tests performed in 2004 have validated the principle of Stokes vector measurements.
Theoretical studies have shown that constraints of $\pm 0.5^{\circ}$ in optical alignment, 
$\pm 0.7^{\circ}$C in temperature and $\pm 4.2\times 10^{-3}$ $V$ in
voltage are sufficient to reach a precision of $1\%$ for the measurement of Stokes parameters,
a precision comparable to the one of Solar System observations.

\section{Conclusions and perspectives}
Numerical simulations based on the core-mantles model 
give first results comparable with polarization phase curves observed in the 
Solar System: negative branch below $20^{\circ}$ and large positive branch with a 
maximum around  $80^{\circ}$. 
The ICAPS microgravity experiments will allow us to validate the numerical studies with new 
optical instruments. On one side, information will be retrieved regarding
the evolution of polarization with particles aggregation under proto-solar conditions,
on the other side wavelength dependence of the polarization will be monitored
and finally the four Stokes parameters of the scattered light should be measured.
First results on cometary studies will be obtained with IPE due to fly on board the ISS in the near future.

\section{Acknowledgements}
We acknowledge support from ESA. 
The development of the experiment is a joint effort and we would like to thank  
the teams involved in the project at Kayser-Threde, EADS, Dutch Space  and Lambda-X
who were extremely helpful regarding the optical validations.



\begin{thebibliography}{999}

\bibitem[Blum et al.(1999)]{jb_mc99b}  Blum, J., Cabane, M., Clowell, J., Giovane, F., Gustafson, B., Henning, T., Holl\"ander, W., Levasseur-Regourd, A.C.,
 Lumme, K., Marshall, J., Muinonen, K., Mukai, T., Nakamura, A., Nuth, J., Piironen, Poppe, T., Prodi, F., Slobodrian, R.J., Vedernikov, A., Wagner, P., 
and Worms, J.C. Interactions in Cosmic and Atmospheric Particle Systems (ICAPS). Experiment proposal to ESA (1999).
\bibitem[Blum et al.(2000)]{jb_jd00} Blum, J., Dorschner, J., El Gorezy, A., Fechtig, H., Feuerbacher, B., Giovane, F.,
 Gustafson, B., Gr\"un, E., Henning, T., Ip, W.H., Keller, H.U., Kempf, S., Klahr, H., Kochan, H.,
 Kozasa, T., Mann, I., Markiewicz, W.J., Metzler, K., Morfill, G., Neuhaus, D., Poppe, T., Ratke, L., 
 Rott, M., Schr\"apler, R., Schwehm, G., Weidenschilling,  S.J., Wurm,  G.  Growth and form of 
planetary seedlings : results from a microgravity aggregation experiment. Phys. Rev. Lett. 85, 2426-2429, 2000.
\bibitem[e.g. Bohren and Huffman(1983)]{cb_dh83} Bohren, C., Huffman, D. Absorption and scattering of light by small particles. 
Wiley, New-York, USA, 1983.
\bibitem[Dorschner et al.(1995)]{jd_bb95} Dorschner, J., Begemann, B., Henning, Th., J\"ager, C., Mutschke, H.
Steps toward interstellar silicate mineralogy: II Study of $Mg-Fe-$silicate
glasses of variable composition. Astron. Astrophys. 300, 503-520, 1995.
\bibitem[Draine and Flatau(2000)]{btd_pjf00} Draine, B.T., Flatau, P.J. User guide for the Discrete Dipole Approximation code DDSCAT 
(Version 5a10),  \underline{http://arxiv.org/abs/astro-ph/0008151v4}, 2000. 
\bibitem[Fulle et al.(2000)]{mf_aclr00} Fulle, M., Levasseur-Regourd, A.C., McBride, M., Hadamcik, E.
In situ dust measurements from within the coma of 1P/Halley: approximation with a dust 
dynamical model. Astron. J. 119, 1968-1977, 2000.
\bibitem[Greenberg and Hage(1990)]{jmg_jih90} Greenberg, J.M., Hage, J.I.  From interstellar dust to comets: a 
unification of observational constraints. Astrophys. J. 361, 260-274, 1990.
\bibitem[Greenberg and Li(1996)]{jmg_al96} Greenberg, J.M., Li, A.  What are the true astronomical silicates ? 
Astron. Astrophys. 309, 258-266, 1996.
\bibitem[see e.g. Greenberg(1997)]{jmg_rev97} Greenberg, J.M. (Ed.) Formation and Evolution of solids in Space. Kluwer, 1997.
\bibitem[Gustafson and Kolokolova(1999)]{basg_lk99} Gustafson, B.A.S., Kolokolova, L. A systematic study of 
light scattering by aggregate particles using the microwave analog technique: angular and wavelength 
dependence of intensity and polarization. J. Geophys. Res.  104, D24, 31711-31720, 1999.
\bibitem[Hadamcik et al.(2002)]{eh_jbr02} Hadamcik,  E., Renard,  J.B., Worms,  J.C., Levasseur-Regourd,  A.C., Masson,  M. 
Polarization of light scattered by fluffy particles (PROGRA$^2$ experiment). Icarus 155, 497-508, 2002.
\bibitem[Hadamcik et al.(2003)]{eh_jbr03} Hadamcik,  E., Renard,  J.B., Levasseur-Regourd, A.C., Worms, J.C.
Laboratory light scattering measurements on "natural" particles with the  PROGRA$^2$ experiment: an overview. 
J. Quant. Spect. Rad. Trans. 79-80, 679-693, 2003.
\bibitem[e.g. Hapke(1993)]{bh93} Hapke, B. Theory of reflectance and emittance spectroscopy. Cambridge Univ. Press, USA, 1993.
\bibitem[Hanner(2003)]{msh03} Hanner, M. S. The scattering properties of cometary dust. J. Quant. Spect. Rad. Trans. 79-80, 695-705, 2003.
\bibitem[Haudebourg(2000)]{vh00} Haudebourg, V. Propri\'et\'es de diffusion lumineuse de particules en suspension: 
transition du r\'egime de Mie \`a celui d'agr\'egats. Utilisation de l'exp\'erience spatiale CODAG/LSU. 
PhD Thesis, Univ. Paris 6, France, 2000. (In French)
\bibitem[Hovenier et al.(2003)]{jwh_hv03} Hovenier, J.W., Volten, H., Mu\~noz, O., van der Zande, W.J., Waters, L.B.F.M.
Laboratory studies of scattering matrices for randomly oriented particles: potentials, problems, and perspectives.
J. Quant. Spect. Rad. Trans. 79-80, 741-755, 2003.
\bibitem[Jessberger et al.(2001)]{ekj_ts01} Jessberger, E.K., Stephan, T., Rost, D., Arndt, P., Maetz, M., 
Stadermann, F.J., Brownlee, D.E., Bradley, J.P., Kurat, G. Properties of interplanetary dust: information 
from collected samples {\sl in} Interplanetary dust, Gr\"un, E., Gustafson, B.A.S., Dermott, S., Fechtig, H. (Eds.) 
Berlin, Springer, 253-294, 2001.
\bibitem[Levasseur-Regourd et al.(2001)]{aclr_vha01} Levasseur-Regourd, A.C., Haudebourg, V., Cabane, M., Worms,  J.C.
Light scattering measurements on dust aggregates, from MASER-8 to ISS. ESA SP. 454, 797-802, 2001. 
\bibitem[Levasseur-Regourd and Hadamcik(2003)]{aclr_eh03}  Levasseur-Regourd, A.C., Hadamcik, E. Light scattering by irregular 
dust particles in the Solar System: observations and interpretation by laboratory measurements. 
J. Quant. Spect. Rad. Trans. 79-80, 903-910, 2003.
\bibitem[Levasseur-Regourd(2004a)]{aclr_sci04} Levasseur-Regourd, A.C. Cometary dust unveiled. Science 304, 1762-1763, 2004a.
\bibitem[Levasseur-Regourd(2004b)]{aclr_yal04} Levasseur-Regourd, A.C. Polarimetry of dust in the Solar System:
remote observations, in-situ measurements and experimental simulations {\sl in} Photopolarimetry in remote sensing. 
NATO Science Series,   Videen, G., Yatskiv, Y., Mishchenko, M. (Eds.) Vol. 161, 393-410, 2004b.
\bibitem[Li and Greenberg(1997)]{li_jmg97} Li, A., Greenberg, J.M.  A unified model of interstellar dust.
Astron. Astrophys. 323, 566-584, 1997.
\bibitem[Meakin(1983)]{pm83} Meakin, P. Formation of fractal clusters and networks by irreversible diffusion-limited 
aggregation. Phys. Rev. Lett. 51, 1119-1122, 1983.
\bibitem[Mishchenko et al.(1999)]{mim_jwh99} Mishchenko, M.I., Hovenier, J.W., Travis, L.D. (Eds.) Light scattering by 
nonspherical particles: theory, measurements, and applications.  Academic Press, San Diego, 1999.
\bibitem[Mu\~noz et al.(2000)]{om_hv00} Mu\~noz, O., Volten, H., de Haan, J.F., Vassen, W., Hovenier, J.W. Experimental determination 
of scattering matrices of Olivine and Allende meteorite particles. Astron. Astrophys. 360, 777-788, 2000.
\bibitem[Petrova et al.(2000)]{evp_kj00} Petrova, E.V., Jockers, K., Kiselev, N.N. Light scattering by 
aggregates with sizes comparable to the wavelength: an application to cometary dust. Icarus 148, 526-536, 2000.
\bibitem[Rietmeijer(1998)]{fr98} Rietmeijer,  F. J. M. Interplanetary dust particles 
{\sl in} Planetary materials. Reviews in mineralogy, Papike, J. J., Ribbe, J. H. (Eds.) Vol. 36, chap. 2, 1998.
\bibitem[Rietmeijer(2002)]{fr02} Rietmeijer,  F. J. M. The earliest chemical dust evolution in the solar nebula. 
Chem. Erde 62, 1-45, 2002.
\bibitem[Tuzzolino et al.(2004)]{ajt_tee04} Tuzzolino, A.J., Economou, T.E., Clark, B.C., Tsou, P., Brownlee, D.E., Green, S.F., 
McDonnell, J.A.M., McBride, N., Colwell, M.T.S.H.
Dust measurements in the coma of comet 81P/Wild 2 by the dust flux monitor instrument. Science 304, 1776-1780, 2004.
\bibitem[Warren(1984)]{sgw84} Warren, S.G. Optical constants of ice from the ultraviolet to the microwave. 
Appl. Opt. 23, 1206-1225, 1984.
\bibitem[West et al.(1997)]{raw_lrd97} West, R.A., Doose, L.R., Eibl, A.M., Tomasko, M.G., Mishchenko, M.I. Laboratory measurements of mineral 
dust scattering phase function and linear polarization. J. Geophys. Res. 102, D14, 16871-16881, 1997.
\bibitem[Worms et al.(1999)]{jcw_jbr99} Worms, J.C., Renard, J.B., Hadamcik, E., Levasseur-Regourd, A.C., Gayet, J.F.
Results of the PROGRA$^2$ experiment: an experimental study in microgravity of scattered polarized light 
by dust particles with large size parameters. Icarus, 142(1), 281-297, 1999. 
\bibitem[Xing and Hanner(1997)]{zx_mh97} Xing, Z., Hanner, M. Light scattering by aggregate particles. 
Astron. Astrophys. 324, 805-820, 1997.
\bibitem[Zolenski et al.(1994)]{ez_tlw94} Zolensky, E., Wilson, T. L., Rietmeijer, F. J. M., Flynn, G. J. (Eds)
Analysis of interplanetary dust. Amer. Inst. Phys. Conf. Proc. vol 310. Amer. Inst. Phys. Press, New York, 1994.


\end{thebibliography}
\end{document}